\documentstyle[referee]{laa}

\begin{document}
\thesaurus{07  
	   (07.09.1; 
	   )}

\title{On Perihelion Concentration of Long-period Comets \\
II. Correct Access}

\author{ Jozef Kla\v{c}ka }
\institute{Institute of Astronomy,
Faculty for Mathematics and Physics,
Comenius University, \\
Mlynsk\'{a} dolina,
842 15 Bratislava,
Slovak Republic}
\date{}
\maketitle

\begin{abstract}
The problem of (non)random distribution of points on sphere
and in space is investigated.
The procedure for obtaining preferred direction (and plane)
for points on the sphere (in the sky) and in the space is discussed.

\end{abstract}

\section{Introduction}
Kla\v{c}ka (1998) has discussed the methods for finding preferred direction
and preferred plane used for several decades in application to perihelion
concentration of long-period comets. The aim of this paper is to present
correct methods for finding preferred direction and preferred plane for
points situated on the sphere, and, also, in general, in the space.

\section{Preferred Direction}

\subsection{Points on Sphere}
Points on the sphere may be characterized
by direction cosines $l_{i}$, $m_{i}$, $n_{i}$. By definition of
direction cosines, the relation
$l_{i}^{2} ~+~m_{i}^{2} ~+~n_{i}^{2} =$ 1 holds.

If we have $N$ points on the sphere, we have a set
of $N$ triples of direction cosines:
$\{ l_{i} \} _{i=1}^{N}$, $\{ m_{i} \} _{i=1}^{N}$, $\{ n_{i} \} _{i=1}^{N}$
(subscript $i$ denotes the $i$-th point).

Let we want to find a preferred direction from the set of $N$ points
situated on the sphere. Formally, we can construct a function of the
preferred direction characterized by the direction cosines
$l$, $m$, $n$, or, $\vec{n_{o}} \equiv$ ( $l$, $m$, $n$):
\begin{equation}\label{1}
\Phi \left ( \vec{n_{o}} \right ) = 2^{- 1} ~ \sum_{i=1}^{N} w_{i} ~
     \left ( \vec{n_{o}} ~-~ \vec{n_{i}} \right ) ^{2} ~-~
     \lambda ~ \left ( \vec{n_{o}} ^{2} ~-~ 1 \right ) ~,
\end{equation}
where $w_{i} ( \ge 0 )$ is the weight of the $i-$th point and
$\lambda$ represents the multiplicator of Lagrange. The preferred
direction $\vec{n_{o}}$ is characterized by the requirement that it minimizes
the function $\Phi \left ( \vec{n_{o}} \right )$. The result is
\begin{eqnarray}\label{2}
l &=& \left ( \sum_{i=1}^{N} w_{i} ~ l_{i} \right ) ~/~ \mu ~,~~
m = \left ( \sum_{i=1}^{N} w_{i} ~ m_{i} \right ) ~/~ \mu ~,~~
n = \left ( \sum_{i=1}^{N} w_{i} ~ n_{i} \right ) ~/~ \mu ~,
\nonumber \\
\mu &=& \sqrt{\left ( \sum_{i=1}^{N} w_{i} ~ l_{i} \right )^{2} ~+~
\left ( \sum_{i=1}^{N} w_{i} ~ m_{i} \right )^{2} ~+~
\left ( \sum_{i=1}^{N} w_{i} ~ n_{i} \right )^{2}} ~.
\end{eqnarray}

Definition of $\Phi \left ( \vec{n_{o}} \right )$ offers also the measure
of spread: $MS$ $\equiv$ $\sqrt{\Phi \left ( \vec{n_{o}} \right )}$ $=$
$\sqrt{1 ~-~ \mu}$, if
$\sum_{i=1}^{N} w_{i} =$ 1 (normalization of weights);
since 0 $\le \mu \le$ 1,
0 $\le MS \le 1$ --
the larger the value of $MS$, the worse determination of the preferred
direction (the case $\mu =$ 0 means that $\vec{n_{o}}$ does not exist).
Thus, in reality, parameter $\mu$ can be used as a measure of the significance
of the existence of the preferred direction.

However, the important question is: What is the motivation for the special
values of the weights $w_{i}$ and of the form of the function
$\Phi \left ( \vec{n_{o}} \right )$? Other forms of
functions $\Phi \left ( \vec{n_{o}} \right )$ would yield other
preferred directions $\vec{n_{o}}$.
The argument in favour of the choice of $\Phi \left ( \vec{n_{o}} \right )$
is in physical equation (4) in section 2.2 of this paper (all points
are situated on the sphere).

\subsection{Points in Space}
We have discussed the determination of the preferred direction $\vec{n_{o}}$
in section 2.1. The procedure was defined by the function
$\Phi ( \vec{n_{o}} )$, which determined also the measure of spread
(we have shown that the quantity $\mu$ may be used, equivalently).
This represents purely geometrical method.

Another method is of a physical character. Using any physical quantity $\vec{f}$
(radius vector, velocity vector, etc.), we can minimize the function
\begin{equation}\label{3}
\Phi ' \left ( \vec{f_{o}} \right ) = \left \{ \sum_{i=1}^{N} w_{i} ~
     \left ( \vec{f_{o}} ~-~ \vec{f_{i}} \right ) ^{2} \right \} ~/~
     \left \{ \sum_{i=1}^{N} w_{i} ~
     \vec{f_{i}} ^{2} \right \} ~.
\end{equation}
The result is
\begin{equation}\label{4}
\vec{f_{o}} = \left \{ \sum_{i=1}^{N} w_{i} ~
	      \vec{f_{i}} \right \} ~/~ \left \{ \sum_{i=1}^{N} w_{i} \right \}  ~,
\end{equation}
and,
\begin{equation}\label{5}
\Phi ' \left ( \vec{f_{o}} \right ) = 1 ~-~ \left \{
       \left ( \sum_{i=1}^{N} w_{i} \right ) ~
       \vec{f_{o}} ^{2} \right \} ~/~
     \left \{ \sum_{i=1}^{N} w_{i} ~ \vec{f_{i}} ^{2} \right \} ~.
\end{equation}
We have 0 $\le \Phi ' \left ( \vec{f_{o}} \right ) \le$ 1 for
$\vec{f_{o}}$ given by Eq. (4). Thus, the quantity
$\Phi ' \left ( \vec{f_{o}} \right )$
may serve as a measure of spread, or, the measure of the significance
of the value $\vec{f_{o}}$.
In the case of kinematical quantities $w_{i} = m_{i}$ (masses) and
$\vec{f_{o}}$ represents the corresponding quantity for the center of mass.

The preferred direction is determined by the relation
$\vec{n ' _{o}}$ $=$ $\vec{f_{o}} ~/~ | \vec{f_{o}} |$.
The better approximation of $\Phi ' \left ( \vec{f_{o}} \right )$
to 1, the less significant is the direction $\vec{n ' _{o}}$.
Of course, $\vec{n ' _{o}}$ is different from $\vec{n_{o}}$ discussed in section
2.1.
In the case
$| \vec{f_{i}} | = f$, $i =$ 1 to $N$, both methods yield the same
results ($\vec{n ' _{o}}$ $=$ $\vec{n_{o}}$).
Approximately, the unit vector
$\left \{ \sum_{i=1}^{N} q_{i} ~ \vec{n_{i}} ~\right \} ~/$
	$\left \{ | \sum_{i=1}^{N} q_{i} ~
	\vec{n_{i}} | \right \}$  -- $q_{i}-$ perihelion of the $i-$th comet --
corresponds to the unit vector
$\left \{ \sum_{i=1}^{N} \vec{n_{i}} ~\right \} ~/$
	$\left \{ | \sum_{i=1}^{N} ~\vec{n_{i}} | \right \}$
if the distribution in $q$ is independent on the observed directions $\vec{n}$.

Eq. (5) may be rewritten to the form
\begin{equation}\label{6}
\Phi ' \left ( \vec{f_{o}} \right ) = \frac{1}{1 ~+~
       \vec{f_{o}} ^{2} ~/~ \vec{\sigma} ^{2}}	~, ~~
\vec{\sigma} ^{2}  \equiv  \sum_{i=1}^{N} w_{i} ~
     \left ( \vec{f_{i}} ~-~ \vec{f_{o}} \right ) ^{2} /
		    \left ( \sum_{i=1}^{N} w_{i} \right )~.
\end{equation}



\subsection{``Local'' Character of the Measure of Spread}
Formulae of the measures of spread discussed in sections 2.1 and 2.2 represent
``global'' character. This means that the quantities $\mu$, $\Phi '$ will yield
a result ``direction is relevant'' already in the case if only one of the
components of $\vec{f_{o}}$, or, one of the spherical angles ( $\alpha$,
$\delta$ ), is very significant. Methods of the sections 2.1 and 2.2 cannot
determine if all components determining preferred direction are significant.

The problem just discussed may be solved by applying central limit theorem.
We calculate for this purpose
\begin{equation}\label{7}
f_{oj} \pm \sigma_{foj} ~,~~~
\sigma_{foj} = \sqrt{\sum_{i=1}^{N} \left ( f_{ij} - f_{oj} \right ) ^{2}
		   / N / \left ( N - 1 \right )} ~,
						    ~~~j = x, y, z
\end{equation}
(analogously for directional cosines -- $f_{ox}$, $f_{oy}$, $f_{oz}$ are
mean values of directional cosines, in this case). Standard tests of
mathematical statistics will decide about relevance of the quantities
$f_{ox}$, $f_{oy}$, $f_{oz}$. Great advantage of this method is that,
in general, not all of the quantities $f_{ox}$, $f_{oy}$, $f_{oz}$
are significant (different from zero). This can help us in better
understanding of the observational data, in their physical and astronomical
interpretation.

\section{Preferred Plane}

\subsection{Points on Sphere}

\subsubsection{Center of the Sphere Lies in the Plane}
Let the preferred plane, characterized by the unit normal vector $\vec{p}$,
is obtained from minimization of the function
\begin{equation}\label{8}
\Theta \left ( \vec{p} \right ) = \sum_{i=1}^{N} w_{i} ~
       ~\left ( \vec{p} \cdot  \vec{n_{i}} \right ) ^{2}
     ~-~  \lambda ~ \left ( \vec{p} ^{2} ~-~ 1 \right ) ~.
\end{equation}
Stationary points are characterized by the equations
$\sum_{i=1}^{N} w_{i} ~
       ~\left ( \vec{p} \cdot  \vec{n_{i}} \right ) ~\vec{n_{i}}
     ~-~  \lambda ~ \vec{p}$ $=$ 0,
$\Theta \left ( \vec{p} \right )$ $=$ $\lambda$ $=$ $\sum_{i=1}^{N} w_{i} ~
       ~\left ( \vec{p} \cdot  \vec{n_{i}} \right ) ^{2} \ge$ 0.
There is minimum of $\Theta \left ( \vec{p} \right )$ for $\lambda (min)$,
maximum for $\lambda (max)$ and there is no extreme of
$\Theta \left ( \vec{p} \right )$ for $\lambda$ corresponding to
$\lambda (min) < \lambda < \lambda (max)$ (the corresponding quadratic forms
are positively definite for $\lambda (min)$, negatively definite for
$\lambda (max)$ and indefinite for the third case -- only two of the
three stationary points correspond to extremes).

Moreover,
\begin{equation}\label{9}
\lambda ( min ) < \left ( 1 ~/~ 3 \right ) ~\sum_{i=1}^{N} w_{i} ~.
\end{equation}
The last condition yields
$\lambda ( min ) < 1 ~/~ 3$ for the case
$\sum_{i=1}^{N} w_{i} =$ 1, and, thus,
$\sum_{i=1}^{N} w_{i} ~
       ~\left ( \vec{p} \cdot  \vec{n_{i}} \right ) ^{2}$
$<$ $1 ~/~ 3$; in other words, the quantity
$\sum_{i=1}^{N} w_{i} ~
       ~\left ( \vec{p} \cdot  \vec{n_{i}} \right ) ^{2}$ $~/~$
$\left ( 1~/~3 \right )$ compares the observed distribution with respect to the
uniform (equal weights) distribution.

The condition
$\lambda ( min ) < \left ( 1 ~/~ 3 \right ) ~\sum_{i=1}^{N} w_{i}$
can be easily proved as follows.
$\Theta \left ( \vec{p} \right )$ $=$ $\sum_{i=1}^{N} w_{i} ~
       ~\left ( \vec{p} \cdot  \vec{n_{i}} \right ) ^{2}$
we write as
$\Theta \left ( \vec{p} \right )$ $=$
$\vec{p} ^{T} ~ \Gamma ~\vec{p}$,
where the tensor $\Gamma$ is defined by its components
$\Gamma _{jk}$ $=$  $\sum_{i=1}^{N} w_{i} ~
       ~\left ( \vec{n_{i}} \right )_{j} ~ \left ( \vec{n_{i}} \right )_{k}$.
The symmetry of $\Gamma$ enables us to find an orthonormal base in which
$\Gamma$ is diagonal (Tr($\Gamma$) is invariant of the transformation --
orthogonal transformation; primed components),
i. e., $\Gamma_{jk}$ $=$  $\sum_{i=1}^{N} w_{i} ~
       ~\left ( \vec{n'_{i}} \right )_{j} ^{2} ~ \delta _{jk}$.
The equation $\Gamma~ \vec{p} = \lambda ~ \vec{p}$ yields for the eigenvalues
$\lambda_{1}$ $=$  $\sum_{i=1}^{N} w_{i} ~
       ~\left ( \vec{n'_{i}} \right )_{1} ^{2}$,
$\lambda_{2}$ $=$  $\sum_{i=1}^{N} w_{i} ~
       ~\left ( \vec{n'_{i}} \right )_{2} ^{2}$,
$\lambda_{3}$ $=$  $\sum_{i=1}^{N} w_{i} ~
       ~\left ( \vec{n'_{i}} \right )_{3} ^{2}$, i. e.,
$\lambda_{1} ~+~\lambda_{2} ~+~ \lambda_{3}$ $=$ $\sum_{i=1}^{N} w_{i}$,
and, thus,
$\lambda (min) \le$ ( 1 / 3 ) $\sum_{i=1}^{N} w_{i}$. The relation
$\lambda_{1} ~+~ \lambda_{2} ~+~ \lambda_{3}$ $=$ $\sum_{i=1}^{N} w_{i}$
may serve as a control in practical calculations.

\subsubsection{General Plane and Sphere}
If the preferred plane is obtained from minimization of the function
\begin{equation}\label{10}
\Theta \left ( \vec{p} , d \right ) = \sum_{i=1}^{N} w_{i} ~
       ~\left ( \vec{p} \cdot  \vec{n_{i}} ~+~ d \right ) ^{2}
     ~-~  \lambda ~ \left ( \vec{p} ^{2} ~-~ 1 \right ) ~,
\end{equation}
then the process of minimization yields for the minimum:
$\sum_{i=1}^{N} w_{i} ~
       ~\left ( \vec{p} \cdot  \vec{n_{i}} ~+~ d \right ) ~\vec{n_{i}}$
     $-$  $\lambda ( min ) ~ \vec{p}$ $=$ 0,
$\sum_{i=1}^{N} w_{i} ~
       ~\left ( \vec{p} \cdot  \vec{n_{i}} ~+~ d \right ) =$ 0,
$\Theta \left ( \vec{p} , d \right )$ $=$ $\lambda (min)$ $=$ $\sum_{i=1}^{N} w_{i} ~
       ~\left ( \vec{p} \cdot  \vec{n_{i}} ~+~ d \right ) ^{2} \ge$ 0.

Moreover, $\lambda (min)$  $<$ $\left ( 1 ~/~ 3 \right )$
$\left \{ \sum_{i=1}^{N} w_{i} ~-~ \left ( \sum_{i=1}^{N} w_{i} \right ) ^{-1}
~ \mu ^{2} \right \}$, where
$\mu = \sqrt{\left ( \sum_{i=1}^{N} w_{i} ~ l_{i} \right )^{2} ~+~
\left ( \sum_{i=1}^{N} w_{i} ~ m_{i} \right )^{2} ~+~
\left ( \sum_{i=1}^{N} w_{i} ~ n_{i} \right )^{2}}$
(the same quantity as in section 2).
The last condition yields
$\lambda (min) < ( 1 ~/~ 3 ) ~ ( 1 ~-~ \mu ^{2} )$ for the case
$\sum_{i=1}^{N} w_{i} =$ 1, and, thus,
$\sum_{i=1}^{N} w_{i} ~
       ~\left ( \vec{p} \cdot  \vec{n_{i}} \right ) ^{2}$
$<$ $( 1 ~/~ 3 ) ~ ( 1 ~-~ \mu ^{2} )$; in other words, the quantity
$\sum_{i=1}^{N} w_{i} ~
       ~\left ( \vec{p} \cdot  \vec{n_{i}} \right ) ^{2}$ $~/~$
$\left ( 1~/~3 \right )$ compares the observed distribution with respect to the
uniform (equal weights; $\mu =$ 0) distribution  -- Eq. (9) may be used.

The condition
$\lambda (min)$  $<$ $\left ( 1 ~/~ 3 \right )$
$\left \{ \sum_{i=1}^{N} w_{i} ~-~ \left ( \sum_{i=1}^{N} w_{i} \right ) ^{-1}
~ \mu ^{2} \right \}$
can be easily proved as follows.
$\Theta \left ( \vec{p} , d \right )$ $=$ $\sum_{i=1}^{N} w_{i} ~
       ~\left ( \vec{p} \cdot  \vec{n_{i}} ~+~ d \right ) ^{2}
     ~-~  \lambda ~ \left ( \vec{p} ^{2} ~-~ 1 \right )$,
we write as
$\Theta \left ( \vec{p} , d \right )$ $=$
$\vec{p} ^{T} ~ \Gamma ~\vec{p}$ $+$ 2 $d~ \vec{p} ^{T} ~ \vec{a}$ $+$
$d^{2} ~ \left ( \sum_{i=1}^{N}  w_{i} \right )$
     $-$  $\lambda ~ \left ( \vec{p} ^{T} ~ \vec{p}  ~-~ 1 \right )$,
where the tensor $\Gamma$ is defined by its components
$\Gamma _{jk}$ $=$  $\sum_{i=1}^{N} w_{i} ~
       ~\left ( \vec{n_{i}} \right )_{j} ~ \left ( \vec{n_{i}} \right )_{k}$,
and the vector $\vec{a} = \sum_{i=1}^{N} \left ( w_{i} ~ \vec{n_{i}} \right )$.
The conditions of stationarity yield
$\Gamma ~\vec{p}$ $+$ $d~ \vec{a}$ $=$ $\lambda ~ \vec{p}$,
$\vec{p} ^{T} ~ \vec{a}$ $+$
$d~ \left ( \sum_{i=1}^{N} w_{i} \right )$ $=$ 0 ~. The last two
equations yield
$\Gamma ~\vec{p}$ $-$ $\left ( \sum_{i=1}^{N} w_{i} \right ) ^{-1}$
$\left ( \vec{p} \cdot \vec{a} \right ) ~\vec{a}$
$=$ $\lambda ~ \vec{p}$, or, in components
$\left \{ \Gamma _{jk} ~-~ \left ( \sum_{i=1}^{N} w_{i} \right ) ^{-1}
~ a_{j} ~ a_{k} \right \} ~ p_{k}$
$=$ $\lambda ~ p_{j}$, or, shortly $\Pi _{jk} ~ p_{k}$ $=$
$\lambda ~ p_{j}$ (summation over repeated indices is supposed in the
last two equations).
The symmetry of $\Pi$ enables us to find an orthonormal base in which
$\Pi$ is diagonal (Tr($\Pi$) is invariant of the transformation --
orthogonal transformation; primed components).
The equation $\Pi~ \vec{p} = \lambda ~ \vec{p}$ yields for the eigenvalues
$\lambda_{j}$ $=$  $\sum_{i=1}^{N} w_{i}
       ~ \left ( \vec{n'_{i}} \right )_{j} ^{2} ~-~
\left ( \sum_{i=1}^{N} w_{i} \right ) ^{-1}$
$\left [ \sum_{i=1}^{N} w_{i} ~ \left ( \vec{n'_{i}} \right )_{j} \right ] ^{2}$,
$j =$ 1, 2, 3, i. e.,
$\lambda_{1} ~+~\lambda_{2} ~+~ \lambda_{3}$ $=$
$\left ( \sum_{i=1}^{N} w_{i} \right )$ $-$
$\left ( \sum_{i=1}^{N} w_{i} \right ) ^{-1}$ $\mu^{2}$, where
$\mu = | \vec{a} |$. Thus,
$\lambda (min)$  $\le$ $\left ( 1 ~/~ 3 \right )$
$\left \{ \sum_{i=1}^{N} w_{i} ~-~ \left ( \sum_{i=1}^{N} w_{i} \right ) ^{-1}
~ \mu ^{2} \right \}$.
The relation for the sum of $\lambda -$s
may serve as a control in practical calculations.

Equation
$\sum_{i=1}^{N} w_{i} ~
       ~\left ( \vec{p} \cdot  \vec{n_{i}} ~+~ d \right ) =$ 0
and the results of the section 2 yield $| d | <$ 1.

\underline{\it Example}: Let us consider
$\left	\{ \vec{n_{i}} \right \} ^{4} _{i=1}$ $=$
$\left \{ \right .$ (1, 0, 1) ~/~ $\sqrt{2}$; ~(0, 1, 1) ~/~ $\sqrt{2}$;
~ ($-$ 1, 0, 1) ~/~ $\sqrt{2}$; ~(0, $-$ 1, 1) ~/~ $\sqrt{2}$ $\left . \right \}$ ;
$w_{i} = 1 / 4$, $i =$ 1 to 4.

The method described in section 2 yields for the preferred direction:
$\vec{n_{o}} =$ (0, 0, 1), $\mu = \sqrt{2} / 2 =$ 0.707.

As for the preferred planes: \\
i) $\Theta \left ( \vec{p} \right ) = \lambda ( min ) =$ 1 / 4,
$\vec{p} =$ ($p_{1}$, $p_{2}$, 0), $p_{1}^{2} ~+~ p_{2}^{2} =$ 1. \\
ii) $\Theta \left ( \vec{p}, d \right ) = \lambda ( min ) =$ 0,
$\vec{p} =$ (0, 0, $(\pm)$ 1), $d = - ~(\pm)~ \sqrt{2} / 2$. \\
Particular results: $\vec{a} = (0, 0, 1/ \sqrt{2})$, $\lambda (min) \le 1 / 6$,
$\Gamma = diag(1/4, 1/4, 1/2)$, $\Pi = diag(1/4, 1/4, 0)$.

\subsection{Points in Space}

\subsubsection{Center of the Coordinate Axes Lies in the Plane}
Let the preferred plane is obtained from minimization of the function
\begin{equation}\label{11}
\Theta \left ( \vec{p} \right ) = \sum_{i=1}^{N} w_{i} ~
       ~\left ( \vec{p} \cdot  \vec{f_{i}} \right ) ^{2}
     ~-~  \lambda ~ \left ( \vec{p} ^{2} ~-~ 1 \right ) ~,
\end{equation}
where $\vec{f_{i}}$ is ``radius'' vector of the $i-$th point in the
(phase-)space.
Stationary points are characterized by the equations
$\sum_{i=1}^{N} w_{i} ~
       ~\left ( \vec{p} \cdot  \vec{f_{i}} \right ) ~\vec{f_{i}}
     ~-~  \lambda ~ \vec{p}$ $=$ 0,
$\Theta \left ( \vec{p} \right )$ $=$ $\lambda$ $=$ $\sum_{i=1}^{N} w_{i} ~
       ~\left ( \vec{p} \cdot  \vec{f_{i}} \right ) ^{2} \ge$ 0.
There is minimum of $\Theta \left ( \vec{p} \right )$ for $\lambda (min)$,
maximum for $\lambda (max)$ and there is no extreme of
$\Theta \left ( \vec{p} \right )$ for $\lambda$ corresponding to
$\lambda (min) < \lambda < \lambda (max)$ (the corresponding quadratic forms
are positively definite for $\lambda (min)$, negatively definite for
$\lambda (max)$ and indefinite for the third case -- only two of the
three stationary points correspond to extremes).

Moreover,
\begin{equation}\label{12}
\lambda \left ( min \right ) < \frac{1}{3} ~ \sum_{i=1}^{N} ~ w_{i}
		       ~\vec{f_{i}} ^{2} ~.
\end{equation}
The last condition yields
$\lambda ( min ) < \sum_{i=1}^{N} ~\vec{f_{i}} ^{2} / ( 3 N )$
for the case $w_{i} = 1 / N$, and, thus,
$\sum_{i=1}^{N} w_{i} ~
       ~\left ( \vec{p} \cdot  \vec{f_{i}} \right ) ^{2}$
$<$ $\sum_{i=1}^{N} ~\vec{f_{i}} ^{2} / ( 3 N )$;
in other words, the quantity
3 $\sum_{i=1}^{N}
       ~\left ( \vec{p} \cdot  \vec{f_{i}} \right ) ^{2}$ $~/~$
$\left \{ \sum_{i=1}^{N} ~\vec{f_{i}} ^{2} \right \}$
compares the observed distribution with respect to the
uniform distribution.

The condition
$\lambda ( min ) < \left ( 1 ~/~ 3 \right )$
$\sum_{i=1}^{N} w_{i} ~\vec{f_{i}} ^{2}$
can be easily proved in an analogous way as it is presented in section 3.1.1.


\subsubsection{General Plane and Space}
If the preferred plane is obtained from minimization of the function
\begin{equation}\label{13}
\Theta \left ( \vec{p} , d \right ) = \sum_{i=1}^{N} w_{i} ~
       ~\left ( \vec{p} \cdot  \vec{f_{i}} ~+~ d \right ) ^{2}
     ~-~  \lambda ~ \left ( \vec{p} ^{2} ~-~ 1 \right ) ~,
\end{equation}
then the process of minimization yields for the minimum:
$\sum_{i=1}^{N} w_{i}$
       $\left ( \vec{p} \cdot  \vec{f_{i}} ~+~ d \right )$ $\vec{f_{i}}$
     $-$  $\lambda ( min ) ~ \vec{p}$ $=$ 0,
$\sum_{i=1}^{N} w_{i} ~
       ~\left ( \vec{p} \cdot  \vec{f_{i}} ~+~ d \right ) =$ 0,
$\Theta \left ( \vec{p} , d \right )$ $=$ $\lambda (min)$ $=$ $\sum_{i=1}^{N} w_{i} ~
       ~\left ( \vec{p} \cdot  \vec{f_{i}} ~+~ d \right ) ^{2}$.

Moreover, $\lambda (min)$  $<$
$\{ \sum_{i=1}^{N} ~w_{i} ~\vec{f_{i}} ^{2}$ $-$
$\left ( \sum_{i=1}^{N} w_{i} \right ) ^{-1}
~ \vec{a} ^{2} \} / 3$, where
$\vec{a} = \sum_{i=1}^{N} \left ( w_{i} ~ \vec{f_{i}} \right )$.
The last condition yields
$\lambda (min)$
$<$ $\{$ [ $ ( 1 / N )$ $\sum_{i=1}^{N}$ $\vec{f_{i}} ^{2}$ ] $-$
$\vec{a} ^{2}$ $\} / 3$
for the case $w_{i} = 1 / N$, and, thus,
$\sum_{i=1}^{N} w_{i}$
       $\left ( \vec{p} \cdot  \vec{f_{i}} ~+~ d \right ) ^{2}$
$<$ $\{$ [ $ ( 1 / N )$ $\sum_{i=1}^{N}$ $\vec{f_{i}} ^{2}$ ] $-$
$\vec{a} ^{2}$ $\} / 3$;
in other words, the quantity
$\sum_{i=1}^{N}
       ~\left ( \vec{p} \cdot  \vec{f_{i}} ~+~ d \right ) ^{2}$ $/$
$\{$ $\left ( 1~/~3 \right )$ $\sum_{i=1}^{N}$ $\vec{f_{i}} ^{2}$
$\}$
compares the observed distribution with respect to the
uniform ($\vec{a} =$ 0) distribution -- Eq. 12 may be used.

The condition
$\lambda (min)$  $<$ $\left ( 1 ~/~ 3 \right )$
$\{ \sum_{i=1}^{N} ~w_{i} ~\vec{f_{i}} ^{2}$ $-$
$\left ( \sum_{i=1}^{N} w_{i} \right ) ^{-1}
~ \vec{a} ^{2} \}$
can be easily proved in a manner analogous to that presented in section 3.1.2.

\section{Simultaneous Consideration of Preferred Direction and Preferred Plane}
We present a general discussion for simultaneous calculation
of the preferred direction and the preferred plane.

Let us consider a function
\begin{eqnarray}\label{14}
\Phi \left ( \vec{n_{o}} , \vec{p} \right ) &=& 2^{- 1} ~
						\sum_{i=1}^{N} w_{i} ~
     \left ( \vec{n_{o}} ~-~ \vec{n_{i}} \right ) ^{2} ~+~
     A~ \sum_{i=1}^{N} w_{i} ~\left ( \vec{p} \cdot  \vec{n_{i}} \right ) ^{2}
     ~-~  \lambda ~ \left ( \vec{n_{o}} ^{2} ~-~ 1 \right )
\\ \nonumber
\nu ~ \left ( \vec{p} \cdot \vec{n_{o}} \right ) ~,
\end{eqnarray}
which is minimized in order to obtain the preferred direction $\vec{n_{o}}$ and the
unit vector $\vec{p}$ normal to the preferred plane containing the center of the
unit sphere; $w_{i} \ge$ 0. Coefficients
$\lambda$, $\rho$, $\nu$ are multiplicators of Lagrange. The coefficient
$A$ ($\ge$ 0) is arbitrary and there is no argument which value of $A$
one should take into account. The case $A =$ 0 corresponds to the case
$\Phi \left ( \vec{n_{o}} \right )$ discussed already in section 2. The case
$\sum_{i=1}^{N} w_{i} ~
     \left ( \vec{n_{o}} ~-~ \vec{n_{i}} \right ) ^{2}$ $\ll$
$A~ \sum_{i=1}^{N} w_{i} ~\left ( \vec{p} \cdot  \vec{n_{i}} \right ) ^{2}$
and $\lambda = \nu =$ 0 corresponds to the case when there is no
apriori relation between $\vec{n_{o}}$ and $\vec{p}$.

We present two simple examples. \\
1. $\left \{ \vec{n_{i}} \right \} ^{3} _{i=1}$ $=$
$\left \{ (1, 0, 0) ; (- 1, 0, 0) ; (0, 0, 1) \right \}$ ;
$w_{i} = 1 / 3$, $i =$ 1 to 3.	\\
It can be easily verified that
$\vec{n_{o}} = (0, 0, 1)$, $\vec{p} = (0, \pm 1, 0)$. Thus,
$\vec{n_{o}} \cdot \vec{p} = 0$. \\
Moreover, in order of presenting some other properties, we offer some other
results:
$\left \{ \sum_{i=1}^{N} w_{i} ~ \left (
\vec{n_{o}} \cdot \vec{n_{i}} \right ) ^{2} \right \}$ $/$
$\left \{ \sum_{i=1}^{N} w_{i} \right \}$ $=$ 1/3;
$\mu$ (defined in section 2) $=$ 1/3, probability that $\mu >$ 0.33
in the case of random uniform distribution is 83 \% ;
$\left \{ \sum_{i=1}^{N} w_{i} ~ \left (
\vec{p} \cdot \vec{n_{i}} \right ) ^{2} \right \}$ $/$
$\left \{ \sum_{i=1}^{N} w_{i} \right \}$ $=$ 0 $= \lambda (min)$;
definition $w_{1} = w_{2} = 1 / ( k + 2 )$, $w_{3} = k / ( k + 2 )$
yields
$\left \{ \sum_{i=1}^{N} w_{i} ~ \left (
\vec{n_{o}} \cdot \vec{n_{i}} \right ) ^{2} \right \}$ $/$
$\left \{ \sum_{i=1}^{N} w_{i} \right \}$ $=$ $k / ( k + 2 )$ and this
may be greater than 1/3 for $k >$ 1, $\mu (k) = k / ( k + 2 )$,
$\lambda (min, k) =$ 0.

\noindent
2. $\left \{ \vec{n_{i}} \right \} ^{5} _{i=1}$ $=$
$\left \{ (1, 0, 0) ; (- 1, 0, 0) ; (0, 1, 0) ; (0, - 1, 0) ; (0, 0, 1)
\right \}$ ;
$w_{i} = 1 / 5$, $i =$ 1 to 5. \\
It can be easily verified that
$\vec{n_{o}} = (0, 0, 1)$, $\vec{p} = (0, 0, \pm 1)$. Thus,
$\vec{n_{o}} \cdot \vec{p} = \pm 1$.

The last example shows that the function
$\Phi \left ( \vec{n_{o}} , \vec{p} \right )$
is reasonable only in the case $\nu \equiv$ 0. Thus, the important property
of Eq. (14) is that minimization of
$\Phi \left ( \vec{n_{o}} , \vec{p} \right )$
yields \underline{independent} equations for
$\vec{n_{o}}$, $\vec{p}$.

\section{Physics and Geometry}
Finally, we show the relation between physics and geometry. We have
$\Theta ' \left ( \vec{p} \right )$ $=$ $\sum_{i=1}^{N} w ' _{i} ~
       ~\left ( \vec{p} \cdot  \vec{f_{i}} \right ) ^{2}$
$=$ $\sum_{i=1}^{N} w ' _{i} ~ | \vec{f_{i}} | ^{2}
       ~\left ( \vec{p} \cdot  \vec{n_{i}} \right ) ^{2}$,
which is equivalent to $\Theta \left ( \vec{p} \right )$ for the case
$w_{i} = w ' _{i} ~ | \vec{f_{i}} | ^{2}$, $i =$ 1 to $N$.
The minimization of
$\Phi ' \left ( \vec{f_{o}} \right )$ $=$ $\sum_{i=1}^{N} w ' _{i} ~
     \left ( \vec{f_{o}} ~-~ \vec{f_{i}} \right ) ^{2}$
yields
$\vec{f_{o}}$ $=$ $\left \{ \sum_{i=1}^{N} w ' _{i} ~
	\vec{f_{i}} \right \} ~/~ \left \{ \sum_{i=1}^{N} w ' _{i} \right \}$
$=$ $\left \{ \sum_{i=1}^{N} w ' _{i} ~ | \vec{f_{i}} | ~
	\vec{n_{i}} \right \} ~/~ \left \{ \sum_{i=1}^{N} w ' _{i} \right \}$ ~,
$\vec{n_{o}}$
$=$ $\left \{ \sum_{i=1}^{N} w ' _{i} ~ | \vec{f_{i}} | ~
	\vec{n_{i}} \right \} ~/$
	$\left \{ | \sum_{i=1}^{N} w ' _{i} ~ | \vec{f_{i}} | ~
	\vec{n_{i}} | \right \}$ ~.
This is equivalent to the minimum of
$\Phi \left ( \vec{n_{o}} \right )$ $=$ $\sum_{i=1}^{N} w_{i} ~
     \left ( \vec{n_{o}} ~-~ \vec{n_{i}} \right ) ^{2}$
for the case
$w_{i} = w ' _{i} ~ | \vec{f_{i}} |$, $i =$ 1 to $N$
(or, $w_{i} = w ' _{i} ~ | \vec{f_{i}} | ~/$
	$\left \{ | \sum_{i=1}^{N} w ' _{i} ~ | \vec{f_{i}} | ~
	\vec{n_{i}} | \right \}$ ).

\section{Conclusion}
We have presented several methods for obtaining preferred direction
and preferred plane. Methods determining significance of the obtained results
are also presented, and, also, tests on uniform distribution can be easily
elaborated using computer modelling.

\acknowledgements
Special thanks to the firm ``Pr\'{\i}strojov\'{a} technika, spol. s r. o.''.
This work was also partially supported by Grants VEGA
No. 1/4304/97 and 1/4303/97.
\end{document}